\documentclass[11pt,twoside]{article}
\usepackage{asp2010}
\usepackage{natbib}

\resetcounters

\begin{document}

\title{The star formation efficiency in Stephan's Quintet intragroup regions}
\author{G. Natale$^1$, R. J. Tuffs$^2$, C. K. Xu$^3$, C. C. Popescu$^1$, J. Fischera$^4$, U. Lisenfeld$^5$, N. Lu$^3$, P. Appleton$^6$, M. Dopita$^7$, P.-A. Duc$^8$, Y. Gao$^9$, W. Reach$^{10}$, J. Sulentic$^{11}$, M. Yun$^{12}$
\affil{$^1$University of Central Lancashire, Preston, PR1 2HE, UK, email: {\tt gnatale@uclan.ac.uk};  
$^2$Max Planck Institute f\"{u}r Kernphysik, Heidelberg, Germany; 
$^3$Infrared Processing and Analysis Center, Caltech, Pasadena, USA; 
$^4$Canadian Institute for Theoretical Astrophysics, University of Toronto, Toronto, Canada; 
$^5$Department de F\'{\i}sica Te\'{o}rica y del Cosmos, Universidad de Granada, Granada, Spain; 
$^6$NASA Herschel Science Center, IPAC, Caltech, Pasadena, USA; 
$^7$Research School of Astronomy \& Astrophysics, The Australian National University, Weston Creek, Australia; 
$^8$Laboratoire AIM, CEA/DSM-CNRS-Universit\'{e} Paris Diderot, Dapnia/Service d'Astrophysique, CEA-Saclay, France; 
$^9$Purple Mountain Observatory, Chinese Academy of Sciences, Nanjing, China; 
$^{10}$Spitzer Science Center, IPAC, Caltech, Pasadena, USA; 
$^{11}$Instituto de Astrof\'{\i}sica de Andaluc\'{\i}a, Granada, Spain; 
$^{12}$Department of Astronomy, University of Massachusetts, Amherst, USA}}

\begin{abstract}
We investigated the star formation efficiency for all the dust emitting sources in Stephan's Quintet. We inferred star formation rates using Spitzer MIR/FIR and GALEX FUV 
data and combined them with gas column density measurements by various authors, in order to position each source
in a Kennicutt-Schmidt diagram. Our results show that the bright IGM star formation regions in SQ present star formation efficiencies consistent with those observed within 
local galaxies. On the other hand, star formation in the intergalactic shock region seems to be rather inhibited.   

\end{abstract}

\section{Introduction}

\begin{figure}[h]
\begin{center}
 \includegraphics[width=4.5in]{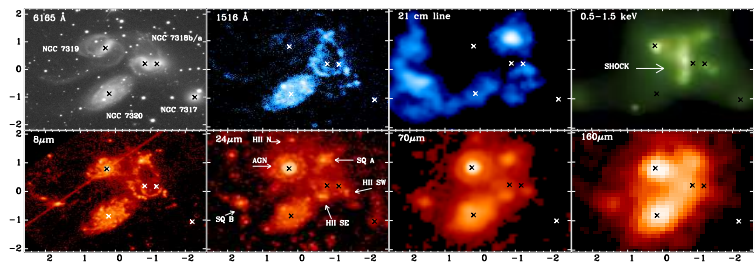} 
 \caption{A multiwavelength view of Stephan's Quintet. Top-panels: SDSS r-band, GALEX FUV, VLA 21 cm line, XMM-NETWON soft X-ray. 
Bottom-panels: Spitzer IRAC $8~\mu m$ and MIPS $24$, $70$ and $160\mu m$. }
   \label{fig1}
\end{center}
\end{figure}

The Stephan's Quintet compact group of galaxies (SQ, $d=94~{\rm Mpc}$) presents widely distributed star formation (SF) in the intergalactic medium (IGM), as clearly observed on the Spitzer 
and GALEX maps of this group (\cite{NA10}, hereafter NA10). 
This SF morphology is usually explained as a consequence of the violent interactions between the group galaxies, which led to stripping of cold gas material
from the main bodies of the galaxies followed by successive triggering of SF in the SQ IGM. In the environment of the IGM the modalities of star formation
could in principle differ from those observed for SF regions within the bodies of galaxies. Therefore it is of particular interest 
to investigate the SF efficiency in IGM SF regions, such as those in SQ. 
NA10 performed an extensive analysis of the Spitzer $8$, $24$, $70$ and $160{\rm \mu m}$ and GALEX FUV data of SQ in order to investigate the origin of the dust 
emission and measure SF rates for each dust emitting source. On the MIR and FIR maps of SQ, shown in Fig.\ref{fig1} together with optical, UV, 21 cm line and soft-X-ray data, several types of sources are detected: 
1) bright compact SF regions located either in the intergalactic medium (SQ-A, SQ-B) or at the periphery of the intruder galaxy NGC 7318b (HII-SE, HII-SW); 2) dust emission from the region of the shock, 
where gas 
has been heated by the collision between NGC 7318b and SQ IGM; 3) an ``extended'' FIR emission component roughly coincident with the X-ray halo. 
One main result of the analysis of these structures has been the discovery of a correlation between the FIR emission and the X-ray emission associated with the IGM sources of the shock region
 and the group halo. A brief description of this result, also considering the possible role of collisionally heated dust in accounting for the FIR - X-ray correlation, 
can be found in \cite{NA12}. 
In this proceeding paper we provide a short summary of the results published by NA10 regarding the 
SF efficiency in SQ, updated to take into account the new CO measurements by \cite{GU12}.          

\section{Star formation rates in SQ}

Star formation rates (SFRs) for all the dust--emitting sources in SQ were estimated using the UV--based SFR calibration from \cite{Salim07}: ${\rm SFR}= 1.08\times10^{-28} F_{\rm UV}~{\rm M_\odot yr}^{-1}$,
where $F_{\rm UV}$ is the UV luminosity density in units of ${\rm erg~s^{-1}Hz^{-1}}$.  
The use of a UV--based SFR calibration implies that our SFR measurements represent averages of the SFR in the last $10^8~{\rm yr}$. By using $H_\alpha$--based SFR 
calibrations instead, one would be able to measure SFRs on scales of $10^7~{\rm yr}$. However, in SQ a substantial fraction of the $H_\alpha$ emission is powered by shocks 
\citep{Xu03}. Therefore, it is not possible to perform a homogeneous set of SFR measurements for all the SQ sources using this kind of calibration (unless a procedure to 
take into account the emission from shock-heated ionized gas is developed and applied to the optical spectra, but this is beyond the scope of this work).  
  On the other hand, morphological comparisons between 
UV and $H_\alpha$ suggest that the UV emission is dominated by young stellar populations. Therefore UV--based SFR calibrations are better suited to estimate SFRs in SQ.\\     
In order to estimate the SFR from the total UV luminosity density $F_{\rm UV}$, using the above formula, we combined measurements from the GALEX UV and Spitzer 
MIR/FIR data, taking into account both dust absorption of UV luminosity reappearing in the FIR and contributions to the FIR luminosity due to optical heating of grains. 
We first performed aperture photometry on the GALEX FUV map to infer the unabsorbed UV emission component. To obtain the absorbed UV emission 
component, we first modeled the observed infrared dust emission source SEDs with a combination of SED templates, to model the dust emission from PDR/HII regions and 
from the diffuse dust in the colder medium surrounding SF regions respectively. The latter template was calculated taking into account stochastic grain heating 
according to the procedure of \cite{Fischera08}. Then we extracted from the best fit SED model the fraction of total dust 
infrared luminosity powered by UV 
photons. Finally we obtained the absorbed component of the luminosity density $F_{\rm UV}$ by simply dividing the UV powered FIR luminosity by a characteristic UV frequency 
width. 
The total SFR we found in SQ is $\approx 7.4~{\rm M_\odot/yr}$, a value consistent with the range of SFRs showed by local galaxies with similar stellar mass of SQ galaxies
\citep{Brinch04}. The measured values of the SFR for each dust emitting source can be found in Table 4 of NA10.       

\section{The Kennicutt-Schmidt plot of SQ sources}
\begin{figure}[h]
\begin{center}
 \includegraphics[width=3.5in]{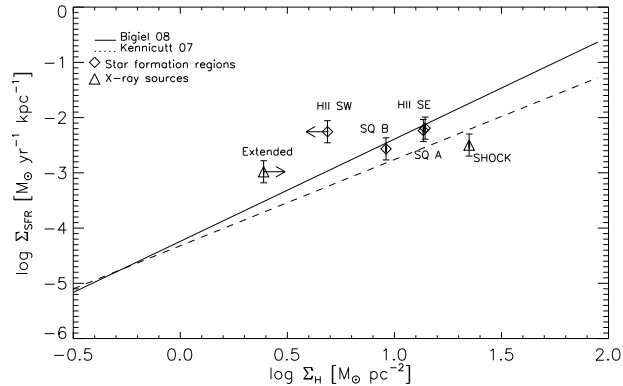} 
 \caption{Kennicutt-Schmidt plot for dust emitting sources in SQ.}
   \label{fig2}
\end{center}
\end{figure}

Fig. \ref{fig2} shows the Kennicutt-Schmidt plot for all the dust--emitting sources in SQ. The plotted SFRs are the values obtained using the method summarized in the 
previous section. The gas densities are the sum of all gas components (ionized, neutral and molecular gas), as obtained from the data published in \cite{Wi02}, 
\cite{Lis02}, \cite{GU12} (GU12, CO measurements, not used in NA10, having a broader bandwidth and covering the shock region, SQ A 
and HII SE) and \cite{T05}. The points on the diagram associated with each
 source are compared
 with empirical KS relations found by \cite{Bigiel08} and \cite{Kennic07} for SF regions within local galaxies. 
Several interesting aspects are evident from this plot. 
First, the two bright intergalactic SF regions SQ A and SQ B present SFRs which can be considered rather ``normal'', that is, consistent with those found for 
local galaxy SF regions. The source HII-SE, a SF region located to the south-east of the intruder galaxy NGC 7318b, presents also a standard SFR. 
This differs from what we found in NA10, because of the inclusion of new CO data from GU12. 
Instead, the source at the south-west of NGC 7318b, HII-SW, seems to be located far away from the area defined by the KS relations for local galaxies, presenting a much
more efficient SFR. For this source, \cite{Lis02} found only an upper limit for the CO column density. There are three possibilities to explain this discrepancy: 
1) a substantial amount of cold gas has been missed by previous observations perhaps because of 
limited bandwidth; 2) there is a large contribution from shocks to the UV emission from this region; 3) the 
discrepancy is real, so this source indeed presents an enhanced SFR whose origin should be investigated.
Another interesting information that can be taken from this plot is that the SF in the region of the shock seems to be rather inhibited. As before for HII-SE, 
also this results differs 
from the one presented in NA10 because of the inclusion of the CO measurements of GU12. Inhibited SF in the shock region has been already claimed by 
several authors, e.g. \cite{A06}, based on the low PAH/warm H2 infrared line ratio, and it is consistent with the highly turbulent gas regime observed in this region (GU12).
The point labelled as ``extended'' in Fig. \ref{fig2} refers to the extended FIR counterpart to the hot diffuse X-ray halo of SQ detected by \cite{T05} 
(see top-right panel of Fig. \ref{fig1}). This point lies above the empirical KS relations apparently suggesting an 
enhanced SFR. However, cold gas masses associated with this dust emission component (extending on scales of $40~{\rm kpc}$) are still to be measured and should be taken into account
to clarify the SFR for this source.       
 
\section{Conclusions and final remarks}
The most firm conclusions from our analysis of the SF efficiency in SQ are the following: 1) the bright IGM SF regions in SQ present star 
formation rates consistent with those observed in local galaxies; 2) the IGM shock region presents rather inhibited SF compared to the average of the
SFR values measured in local galaxies. In NA10 we found that widely distributed 
SF is the powering radiation source for the extended dust emission we detected, apparently associated with the X-ray halo of the group. This component of star 
formation in SQ can be due to SF in tidal debris but also to IGM SF fueled by cooling of hot halo gas. The latter explanation, in effect a 
mechanism for warm accretion out of the IGM, would account for the correlation between the FIR and X-ray emission found by NA10.  
   

\begin{thebibliography}{}

\bibitem[Appleton et al.(2006)]{A06}
{Appleton, P. N., Xu, K. C., Reach, W., Dopita, M. A. et al.} 2006, \textit{ApJ},639,51

\bibitem[Bigiel et al.(2008)]{Bigiel08}
{Bigiel, F., Leroy, A., Walter, F., Brinks, E. et al.} 2008, \textit{AJ},136,2846

\bibitem[Brinchmann et al.(2004)]{Brinch04}
{Brinchmann, J., Charlot, S., White, S. D. M., Tremonti, C.} 2004, \textit{MNRAS},351,1151

\bibitem[Fischera et al.(2008)]{Fischera08}
{Fischera, J., Dopita, M. A.} 2008, \textit{ApJS},176,164

\bibitem[Guillard et al.(2012)]{GU12}
{Guillard et al.} 2012, \textit{ApJ}, submitted

\bibitem[Kennicutt et al.(2007)]{Kennic07}   
{Kennicutt, R. C., Calzetti, D., Walter, F., Helou, G. et al.} 2007, \textit{ApJ},671,333

\bibitem[Lisenfeld et al.(2002)]{Lis02}      
{Lisenfeld, U., Braine, J., Duc, P.-A., Leon, S. et al.} 2002, \textit{A\&A},394,823

\bibitem[Natale et al.(2010)]{NA10}
{Natale, G., Tuffs, R. J., Xu, C. K., Popescu, C. C., et al. } 2010,
\textit{ApJ},725,955

\bibitem[Natale et al.(2012)]{NA12}
{Natale, G. et al.} 2012, Proceedings IAU Symp. 284, ``The Spectral Energy Distribution of Galaxies'', eds. R.J. Tuffs and C.C Popescu,
Cambridge University Press

\bibitem[Salim et al.(2007)]{Salim07}
{Salim, S., Rich, R. M., Charlot, S., Brinchmann, J.} 2007, \textit{ApJS},173,267
	
\bibitem[Trinchieri et al.(2005)]{T05}
{Trinchieri, G., Sulentic, J., Pietsch, W., Breitschwerdt, D.} 2005, \textit{A\&A},444,697

\bibitem[Williams et al.(2002)]{Wi02}
{Williams, B. A., Yun, M. S., Verdes-Montenegro, L.} 2002, \textit{AJ}, 123, 2417

\bibitem[Xu et al.(2003)]{Xu03}
{Xu, C. K., Lu, N., Condon, J. J., Dopita, M., Tuffs, R. J.} 2003, \textit{ApJ}, 595, 665


\end{thebibliography}

\end{document}